\documentclass[a4paper, amsfonts, amssymb, amsmath, reprint, showkeys, nofootinbib, twoside]{revtex4-1}
\usepackage[english]{babel}
\usepackage[utf8]{inputenc}
\usepackage[colorinlistoftodos, color=green!40, prependcaption]{todonotes}
\usepackage{amsthm}
\usepackage{mathtools}
\usepackage{physics}
\usepackage{xcolor}
\usepackage{graphicx}
\usepackage[left=23mm,right=13mm,top=35mm,columnsep=15pt]{geometry} 
\usepackage{adjustbox}
\usepackage{placeins}
\usepackage[T1]{fontenc}
\usepackage{lipsum}
\usepackage{csquotes}

\begin{document}
\title{Carrier Transport in Electrically-Driven \\ Photonic Crystal Membrane Lasers}

\author{Mathias Marchal, Evangelos Dimopoulos, Kasper Spiegelhauer, Nikolaos Chatzaras, Marco Saldutti, Kresten Yvind, Yi Yu, Jesper Mørk*}
    \email[Correspondence email address: ]{jesm@dtu.dk}
    \affiliation{DTU Electro, Technical University of Denmark, DK-2800 Kgs. Lyngby, Denmark}
    \affiliation{NanoPhoton - Center for Nanophotonics, Technical University of Denmark, DK-2800 Kgs. Lyngby, Denmark}

\date{\today} 


\begin{abstract}
We model carrier transport in photonic crystal lasers with lateral current injection through two-dimensional (2D) finite-volume simulations. Though such lasers can achieve ultra-low threshold currents, leakage paths reduce the carrier injection efficiency. The design is evaluated through its performance in terms of injection efficiency, internal quantum efficiency, and IV characteristics. Our model predicts the presence of unconventional leakage paths, explaining experimental observations of low injection efficiencies and enhanced spontaneous recombination at doping interfaces. Carrier leakage paths arise due to insufficient injection of holes into the active region, leading to an electric field that increases the energy barrier for electrons, thereby reducing the injection efficiency. The spatial profile of the p-doped region is shown to play a critical role in achieving a high electrical injection efficiency and low-threshold lasing. The model is an important step towards modelling and optimizing properties of 2D photonic crystal membrane lasers.
\end{abstract}


\maketitle

\section{\label{sec:level1}Introduction}
From the first demonstration of the semiconductor laser in 1962 \cite{Marshall1962, Hall1962}, the field of semiconductor lasers has seen significant advancements in miniaturization and performance. Semiconductor lasers have widespread applications ranging from optical communication and sensing to LIDARs, optical imaging, material processing, etc. A crucial milestone in the development of efficient semiconductor lasers was the invention of the buried heterostructure.\cite{Hayashi1970} In a buried heterostructure (BH), the active medium, often consisting of quantum wells, is buried inside a higher bandgap material. The potential well created by the buried heterostructure efficiently captures and confines carriers resulting in more efficient lasers. The conventional vertical current injection (VCI) scheme features a p-i-n structure where the doped regions are placed above and below the buried heterostructure. Current blocking mechanisms (e.g. p-n-p-n structures) can be employed at either side of the active region to reduce carrier leakage.\cite{Kim2004, Elattar_2023} This configuration results in efficient carrier injection. \\ \\ In the past two decades, photonic crystal (PhC) lasers have seen tremendous progress in the reduction of energy consumption and integration on the Si platform.\cite{Ellis2011,Matsuo2013,Matsuo2019, Takeda2021, Dimopoulos2023,Crosnier2017, Mork2024} Often, a lateral current injection (LCI) scheme, illustrated in Figure \ref{fig:Schematics}(a), is employed. The lateral current injection scheme offers greater flexibility in the fabrication process and allows one to shape the doping profile, reducing doping-induced losses. With a vertical current injection scheme, on the other hand, the doping profile cannot be accurately controlled, leading to high doping-induced losses which limit the Q-factor and threshold current.\cite{Crosnier2017} The combination of a lateral current injection scheme with a new generation of nano-scale buried heterostructures has led to the demonstration of the first sub-$\mu$A continuous wave room temperature semiconductor laser.\cite{Dimopoulos2023} However, despite the low threshold current, the design has been shown to suffer from new leakage paths, limiting the device's injection efficiency.\cite{Dimopoulos2022} The lateral current injection scheme does not allow for current blocking mechanisms used in the vertical current injection lasers. A different approach, proposed by Matsuo et al. \cite{Matsuo2019}, introduces a current blocking trench by etching a trench in the PhC slab to eliminate carrier leakage paths. Understanding and modeling the electrical properties of LCI PhC lasers is crucial to further improve their performance. \\ \\ Previous work by Sargent et al.\cite{Sargent1998} has highlighted the importance of strong carrier confinement required to suppress leakage paths. Berrier et al. \cite{Berrier2007} characterized n-doped InP PhC structures and the effects of dry etching induced modifications on the electrical properties of the PhC holes. More recently, Petykiewicz et al. \cite{Petykiewicz2012} studied laterally doped PhC GaAs p-i-n diode structures and emphasized the role of the tapered doping profile in reducing carrier leakage. Despite these efforts to reduce carrier leakage in lateral current injection PhC lasers, carrier leakage paths remain present.  \\ \\ In this work, we study carrier transport in InP photonic crystal membrane lasers through two-dimensional (2D) simulations. The model explains experimental observations of enhanced spontaneous recombination at the interfaces of the p-doped region and the reduced injection efficiency into the active region. In Section \ref{sec:DevOverview}, we give a brief overview of the device structure of the device under investigation. In Section \ref{sec:ModelOver}, we give a detailed overview of the model and material parameters used. Section \ref{sec:Results}, contains a discussion of both simulated and experimental results. We discuss the observed trends in injection efficiency and internal quantum efficiency, the presence of carrier leakage paths and their origin, and the influence of the offset of the p-doped region on the leakage. Our simulations explain experimental observations of enhanced spontaneous recombination near the p-i InP interface.

\section{Device overview}\label{sec:DevOverview}
\begin{figure}[]
    \includegraphics[width=0.9\linewidth]{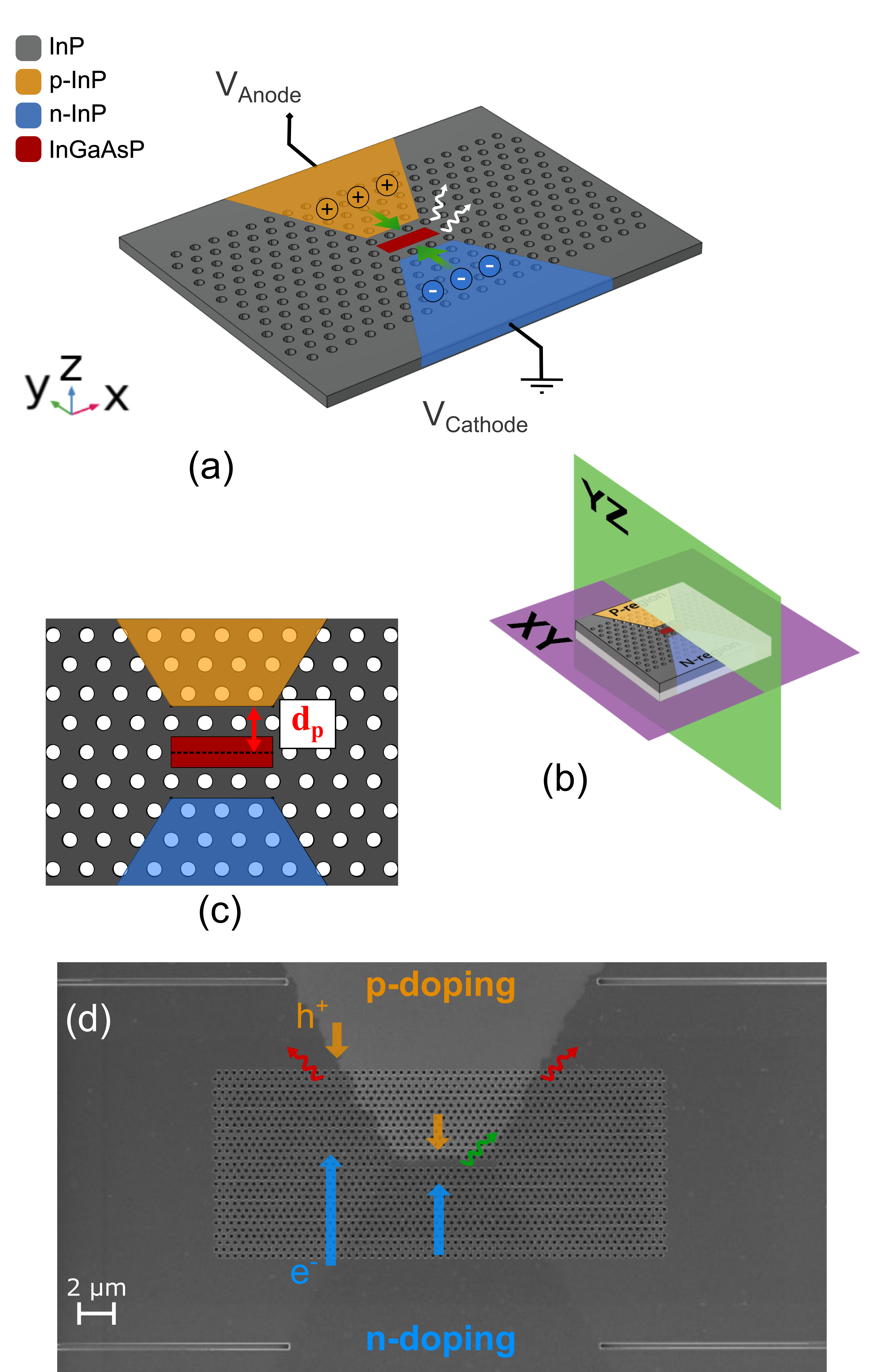}
    \caption{\label{fig:Schematics}(a) Schematic of an L3 line-defect laser. The red shaded area marks the position of the BH. (b) Schematic showing the XY and YX simulation planes. (c) Schematic indicating the doping offset of the p-doped region. (d) Top-view SEM image of an L12 laser. The blue and orange arrows indicate the electron and hole current, respectively. The emission from the BH and p-i InP interface are depicted by the green and red arrows, respectively. }
\end{figure}

Figure \ref{fig:Schematics}(a) shows a schematic of the 2D photonic crystal membrane laser under investigation. The design features an L$N$ line-defect cavity created by omitting $N$ holes in an otherwise periodic triangular lattice of air holes in an InP slab. All simulations are performed on an L$3$ cavity, where three holes were omitted to form the cavity. The lattice constant, $a$, and hole radius, $r$, are 440 nm and 120 nm, respectively. The doping profile is tapered to match the cavity length and reduce leakage. The device consists of a 250nm thick InP layer directly bonded on a SiO$_2$/Si wafer. More details on the fabrication process can be found in Ref. \cite{Dimopoulos2022}. The buried heterostructure active region comprises a single InGaAsP/InAlGaAs quantum well (QW) with a length equal to the defect's length and a width of 400 nm. A lateral p-i-n junction is employed to facilitate carrier injection. Due to the large size of the laser devices and the related computational and complexity limitations of the software, all simulations performed are two-dimensional simulations. Figure \ref{fig:Schematics}(b) shows the different 2D simulation planes that were studied and will be referred to as the XY and YZ simulations. The p-doping offset, $d_\mathrm{p}$, is defined as the distance from the center of the BH to the edge of the p-doped region, as shown in Figure \ref{fig:Schematics}(c). A scanning electron microscope (SEM) image of an L12 line-defect photonic crystal laser is shown in Figure \ref{fig:Schematics}(d). The arrows indicate the flow of electrons (blue) and holes (orange), and the emission from the active region (green) and the p-i InP interface (red). Unless stated otherwise, simulations are performed on the L3 design in which the BH length and doping profiles match the cavity length. The offset of the n- and p-doped regions to the center of the cavity is 1.5 $a_\mathrm{y}$ with $a_\mathrm{y}$ = $\frac{\sqrt{3}}{2}a$. 

\section{Simulation of carrier transport in PhC membrane lasers}
\subsection{Model overview} \label{sec:ModelOver}
We model carrier transport by solving carrier drift-diffusion equations self-consistently with Poisson's equation. The carrier drift-diffusion equations,\cite{Liu_2005} given by Equation \eqref{eq:Jn} and \eqref{eq:Jp}, are solved to obtain the spatial dependency of the density of the electrons, $n$, and the holes, $p$. The subscripts n and p indicate that the variable is related to electrons and holes, respectively. 
\begin{equation}  
    J_\mathrm{n} = q \mu_\mathrm{n} n E + q D_\mathrm{n} \Delta n
    \label{eq:Jn}
\end{equation}
\begin{equation}
    J_\mathrm{p} = q \mu_\mathrm{p} p E - q D_\mathrm{p} \Delta p
    \label{eq:Jp}
\end{equation} 
Further, $J_\mathrm{n,p}$ are the electron and hole current density, $D_\mathrm{n,p}$ are the diffusion coefficients for electrons and holes, and $\mu_\mathrm{n,p}$ are the electron and hole mobilities, respectively. In the right-hand side of Equation \eqref{eq:Jn} and \eqref{eq:Jp}, the first term is the drift current caused by the presence of an electric field, while the second term is a diffusion current due to a gradient in carrier concentration. To take charge conservation into account, these equations are complemented by the auxiliary continuity equations given by \cite{Liu_2005}:
\begin{equation}
    \frac{\partial n}{\partial t} = \frac{1}{q}\nabla \cdot J_\mathrm{n} - U_\mathrm{n} - \sigma_\mathrm{a}R_\mathrm{st}
    \label{eq:Contn}
\end{equation}
\begin{equation}
    \frac{\partial p}{\partial t} = \frac{1}{q}\nabla \cdot J_\mathrm{p} - U_\mathrm{p} - \sigma_\mathrm{a}R_\mathrm{st}
    \label{eq:Contp}
\end{equation}
where $U_\mathrm{n,p}$ is the total recombination rate for electrons and holes, respectively. The total generation rate is assumed to be zero, and the total recombination rate consists of the spontaneous recombination rate, $R_\mathrm{sp}$, the bulk and surface Shockley-Reid-Hall (SRH) recombination rate, $R_\mathrm{SRH}$, and the Auger recombination rate, $R_\mathrm{Auger}$. Bulk and surface SRH recombination are included using the Shockley-Reid-Hall model, which assumes a defect state at the midgap energy. The trap-assisted surface recombination is included as a boundary condition on all etched surfaces. As there are no dry-etched surfaces present in the YZ simulation plane, surface effects are not accounted for in the YZ simulations. Furthermore, the net stimulated recombination rate, $R_\mathrm{st}$, inside the BH, is explicitly added. This is emphasized by the factor $\sigma_\mathrm{a}$, which is 1 inside the BH region and 0 outside. More details on the models used for the recombination mechanisms can be found in Note S1 in the Supporting Information of this paper.\\ \\ Additionally, Poisson's equation \cite{Liu_2005} is solved to find the electrostatic potential due to the free carrier and ionized impurity concentrations;
\begin{equation}
    \nabla \cdot [-\nabla\Phi] = \frac{q}{\epsilon_\mathrm{0}\epsilon_\mathrm{r}}[p-n+N_\mathrm{D}^+ - N_\mathrm{A}^-]
    \label{eq:Poisson}
\end{equation}
Here, $\Phi$ is the electrostatic potential, $q$ is the elementary charge, $\epsilon_\mathrm{r}$ and $\epsilon_\mathrm{0}$ are the relative permittivity and dielectric constant, respectively, and $N_\mathrm{D}^+ - N_\mathrm{A}^-$ is the total charge due to ionized impurities. Together, Equation \eqref{eq:Jn}-\eqref{eq:Poisson} describe the transport problem for semiconductors.\\ \\ 
Stimulated recombination is included to capture the lasing dynamics of our device. The stimulated recombination rate is defined as \cite{Coldren_Corzine_ch2}:
\begin{equation}
    R_{\mathrm{st}} = v_\mathrm{g} g(\mathbf{r}) N_\mathrm{p}
    \label{eq:Rstim}
\end{equation}
Here, $v_\mathrm{g} = c/n$ is the group velocity, $g(\mathbf{r})$ the position-dependent gain with the vector $\mathbf{r}$ the position inside the BH, and $N_\mathrm{p}$ the photon density. \\ \\The photon density is found by solving the steady-state form of the photon density rate equation, given by Equation \eqref{eq:Np} and based on the model proposed in Ref. \cite{Coldren_Corzine_ch2}, at each voltage step. 
\begin{equation}
    \frac{dN_\mathrm{p}}{dt} = [\Gamma v_\mathrm{g} \overline{g}_\mathrm{BH} - \frac{1}{\tau_\mathrm{p}}]N_\mathrm{p} + \Gamma \beta_\mathrm{sp}\overline{R}_\mathrm{sp,BH}
    \label{eq:Np} 
\end{equation}
While the electron and hole density in Equation \eqref{eq:Contn} and \eqref{eq:Contp} are position dependent, the photon density, $N_\mathrm{p}$, in Equation \eqref{eq:Np} is independent of position. This is the result of the assumption that there is only one dominant mode whose field profile variation is negligible inside the BH. This is incorporated in the model by calculating the photon density using the integrated gain, $\overline{g}_\mathrm{BH}$, and integrated spontaneous recombination, $\overline{R}_\mathrm{sp,BH}$, where both are integrated over the BH active region. Furthermore, $\Gamma$ is the confinement factor, $\tau_\mathrm{p}$ the photon lifetime, and $\beta_\mathrm{sp}$ the so-called beta-factor, defined as the fraction of spontaneously emitted photons coupling to the lasing mode.\\ \\
The position-dependent gain inside the active region is defined as \cite{Coldren_Corzine_ch4}: 
\begin{equation}
    g(\textbf{r}) = g_\mathrm{0} \cdot (f_\mathrm{2}(\mathbf{r}) - f_\mathrm{1}(\mathbf{r}))
    \label{eq:gain}
\end{equation}
where the maximum gain, $g_\mathrm{0}$, is a material parameter and set to 2000 cm$^{-1}$ in line with values found in literature,\cite{Matsuo2013} and $f_\mathrm{1}$ and $f_\mathrm{2}$ are the valence and conduction bands' occupation probabilities, respectively. InGaAsP is used as active material in the simulations and provides the gain. \\ \\
In summary, our model solves Equation \eqref{eq:Jn}-\eqref{eq:gain} in a self-consistent manner. It includes carrier transport through Equation \eqref{eq:Jn}-\eqref{eq:Poisson} and accounts for gain and stimulated emission through Equation \eqref{eq:Rstim} and \eqref{eq:gain}, making it suitable to study carrier transport in semiconductor nanolasers.\\ \\
To account for the possibility of carriers escaping the active region after being injected into it, we define the injection efficiency as the fraction of the total current that results from recombinations in the active region and is given by:
\begin{equation}
    \eta_{\mathrm{inj}} = \frac{\int_{V_\mathrm{a}} \sum_\mathrm{m} R_\mathrm{m} dV}{I_{\mathrm{tot}}/q}
    \label{eq:IE}
\end{equation}
Here, $V_\mathrm{a}$ is the volume of the active region, $R_\mathrm{m}$ represents each of the recombination rates included in Equation \eqref{eq:Contn} and \eqref{eq:Contp}, $I_{\mathrm{tot}}$ is the total current flowing between the anode and cathode and $q$ the elementary charge.\\ \\  Another important figure of merit is the internal quantum efficiency, $\eta_{\mathrm{IQE}}$. It is defined as the fraction of the recombination events occurring in the active region that are radiative recombination events and is given by:
\begin{equation}
    \eta_{\mathrm{IQE}} = \int_{V_\mathrm{a}}\frac{R_\mathrm{sp.}+R_\mathrm{st.}}{\sum_m R_\mathrm{m}}dV
    \label{eq:IQE}
\end{equation}
The total efficiency, $\eta_{\mathrm{tot}}$, with which injected carriers result in radiative recombination events can now be defined as:
\begin{equation}
    \eta_{\mathrm{tot}} = \eta_{\mathrm{inj}} \cdot \eta_{\mathrm{IQE}}
\end{equation}\\ \\
We implement our model using COMSOL's Semiconductor Module,\cite{COMSOL} supplemented by stimulated emission according to Equation \eqref{eq:Rstim},\eqref{eq:gain}. Due to the 2D nature of the simulations, metal contact boundary conditions are applied at the outer edge of the p- and n-doped regions and are modeled as ideal ohmic contacts. In COMSOL, materials can only be modeled as bulk, and therefore, the quantum well BH region cannot be accurately modeled in the simulations. However, in order to improve the carrier density dependence of the material gain, we include quantization effects when calculating the density of states and gain. It was verified by additional simulations that these approximations only affect the quantitative behaviour of the model and not the observed trends and, therefore, the results and discussion given below remain valid. More details on the material model used for the BH active region and the additional simulations performed can be found in Note S2 in the Supporting Information. \\ \\
The electron and hole mobilities in InP are $\mu_\mathrm{n}$ = 3500 cm$^2$(Vs)$^{-1}$ and $\mu_\mathrm{p}$ = 200 cm$^2$(Vs)$^{-1}$, respectively.\cite{Sargent1998} The p- and n-doping concentrations are $N_\mathrm{D}$ = 1$\cdot$10$^{18}$ cm$^{-3}$ and $N_\mathrm{A}$ = 3$\cdot$10$^{18}$ cm$^{-3}$, while a background n-doping concentration of $N_\mathrm{bg}$ = 1$\cdot$10$^{15}$ cm$^{-3}$ was assumed. These values are in accordance with the doping parameters of fabricated samples.\cite{Marchevsky2019} The doping concentrations are uniform in the doped regions, and complete ionization of the dopants is assumed. The surface recombination velocity was set to 1$\cdot 10^4$ cm s$^{-1}$ in line with values found in literature.\cite{Yu2013_v2} Note S3 in the Supporting Information contains an overview of the used material parameters. The results of the mesh convergence study performed are included in Note S4 of the Supporting Information. 
\subsection{Simulation results}  \label{sec:Results}
\begin{figure}[ht]
    \includegraphics[width=0.8\linewidth]{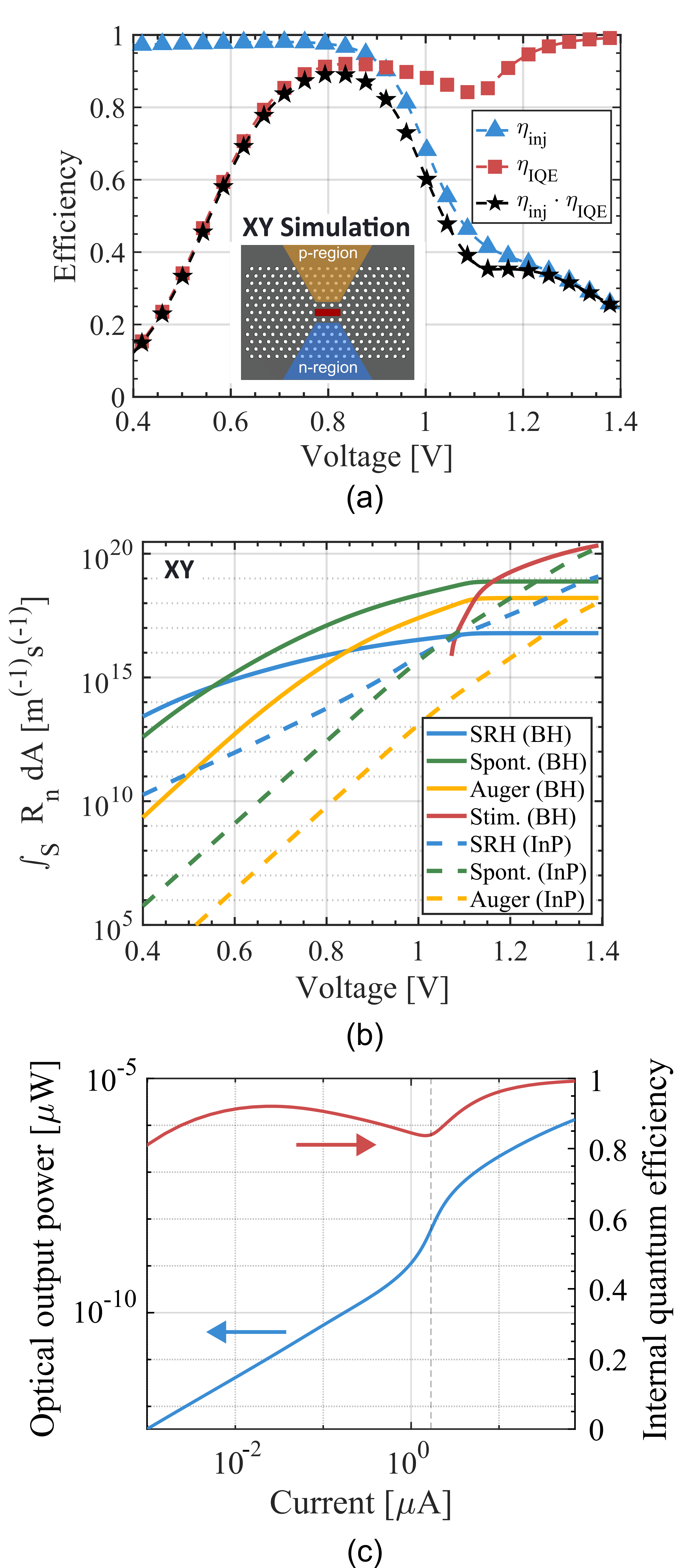}
    \caption{\label{fig:XY_IE_IQE_IV}~(a) Injection efficiency, internal quantum efficiency, and total efficiency as a function of the applied voltage for the XY simulation. (b) Recombination rates per unit volume integrated over the BH region (solid lines) and the background InP region (dashed lines) versus the applied voltage for the XY simulation. (c) Simulated output power and internal quantum efficiency versus applied current. The dashed line indicates the laser threshold and is calculated as the maximum of the first derivative of the log(L)-log(I) curve.}
\end{figure}
\begin{figure}[h]
    \includegraphics[width=0.85\linewidth]{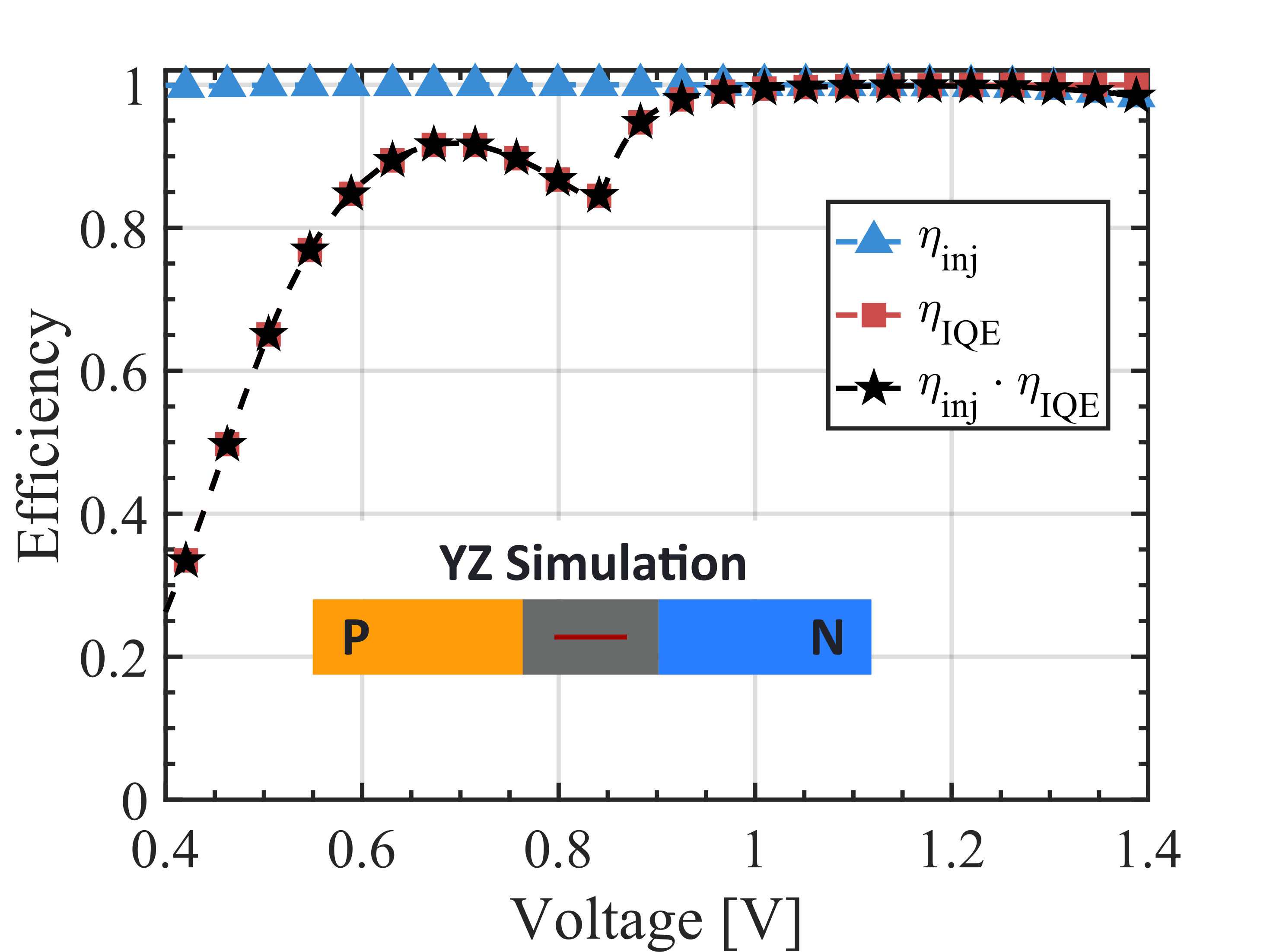}
    \caption{\label{fig:YZ_IE_IQE_IV} Injection efficiency, internal quantum efficiency, and total efficiency as a function of the applied voltage for the YZ simulation}
\end{figure}
\begin{figure}[h]
    \centering
    \includegraphics[width=1\linewidth]{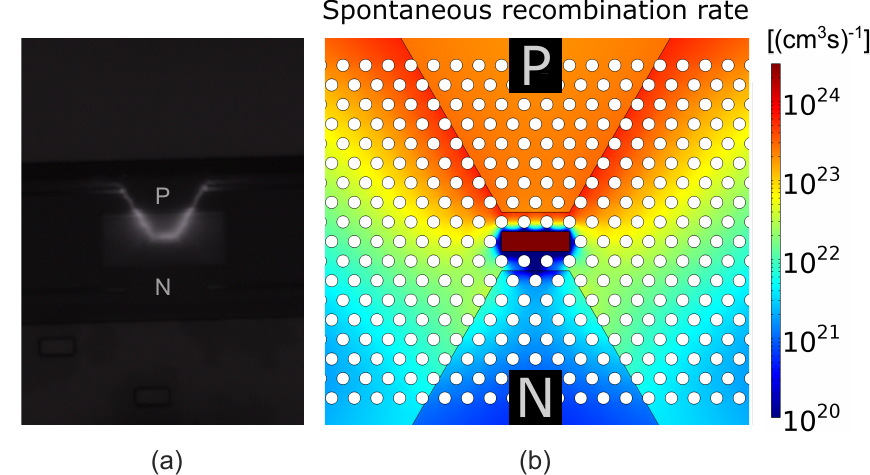}
    \caption{\label{fig:SpontRec} (a) Dimmed microscope image of an L12 line-defect laser at forward bias captured with a Si camera showing enhanced spontaneous recombination in InP at the p-i interface. (b) Surface plot of the simulated spontaneous recombination rate at an applied voltage $V_{\mathrm{bias}}$ = 1.3 V for the XY-plane.}
\end{figure}
\begin{figure}[h]
    \centering
    \includegraphics[width=1\linewidth]{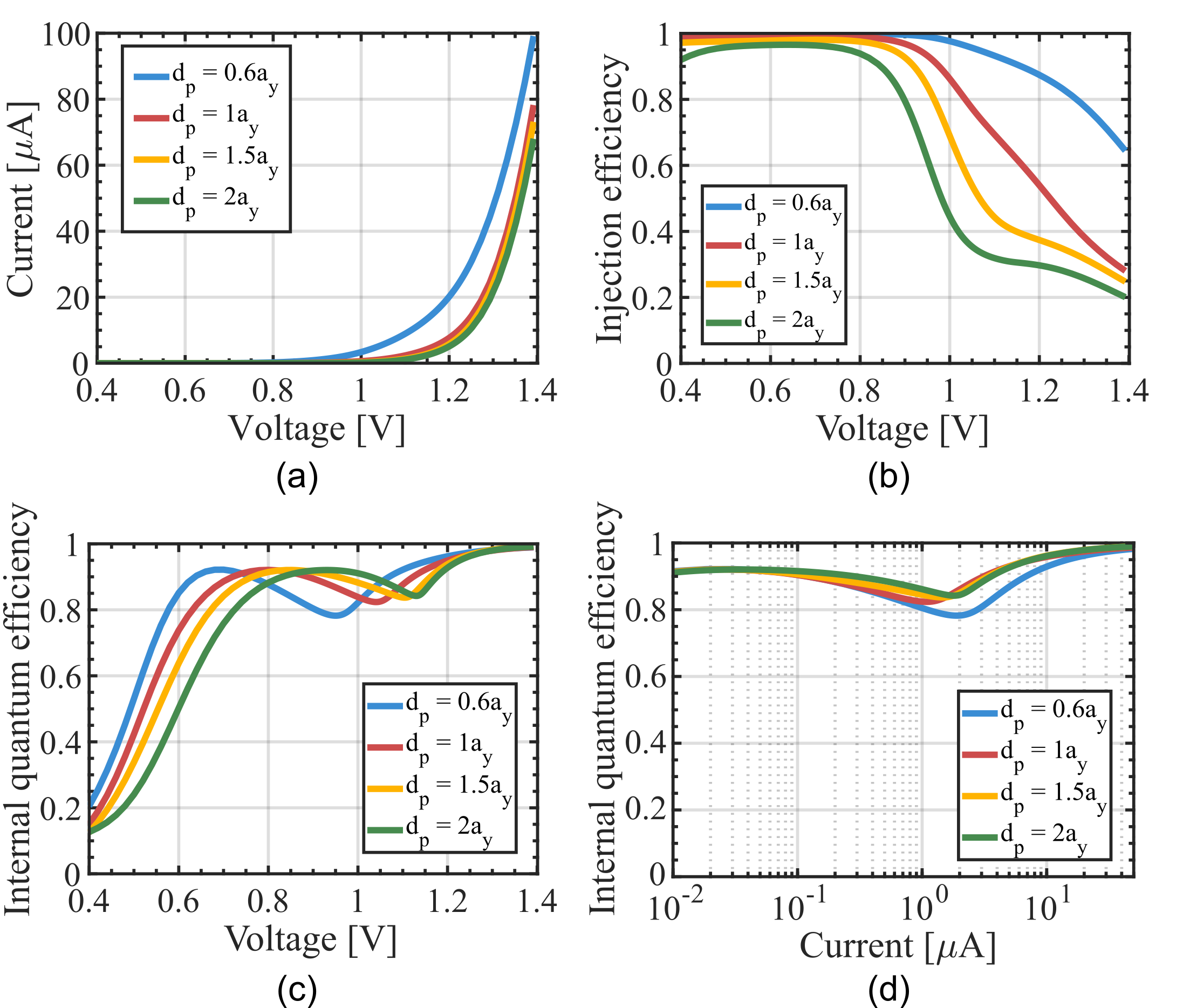}
    \caption{\label{fig:pdoping} Current (a), injection efficiency (b), and internal quantum efficiency (c) for simulations with varying p-doping offset versus applied bias. (d) Internal quantum efficiency versus current.}
\end{figure}
\begin{figure}[t]
    \centering
    \includegraphics[width=0.9\linewidth]
    {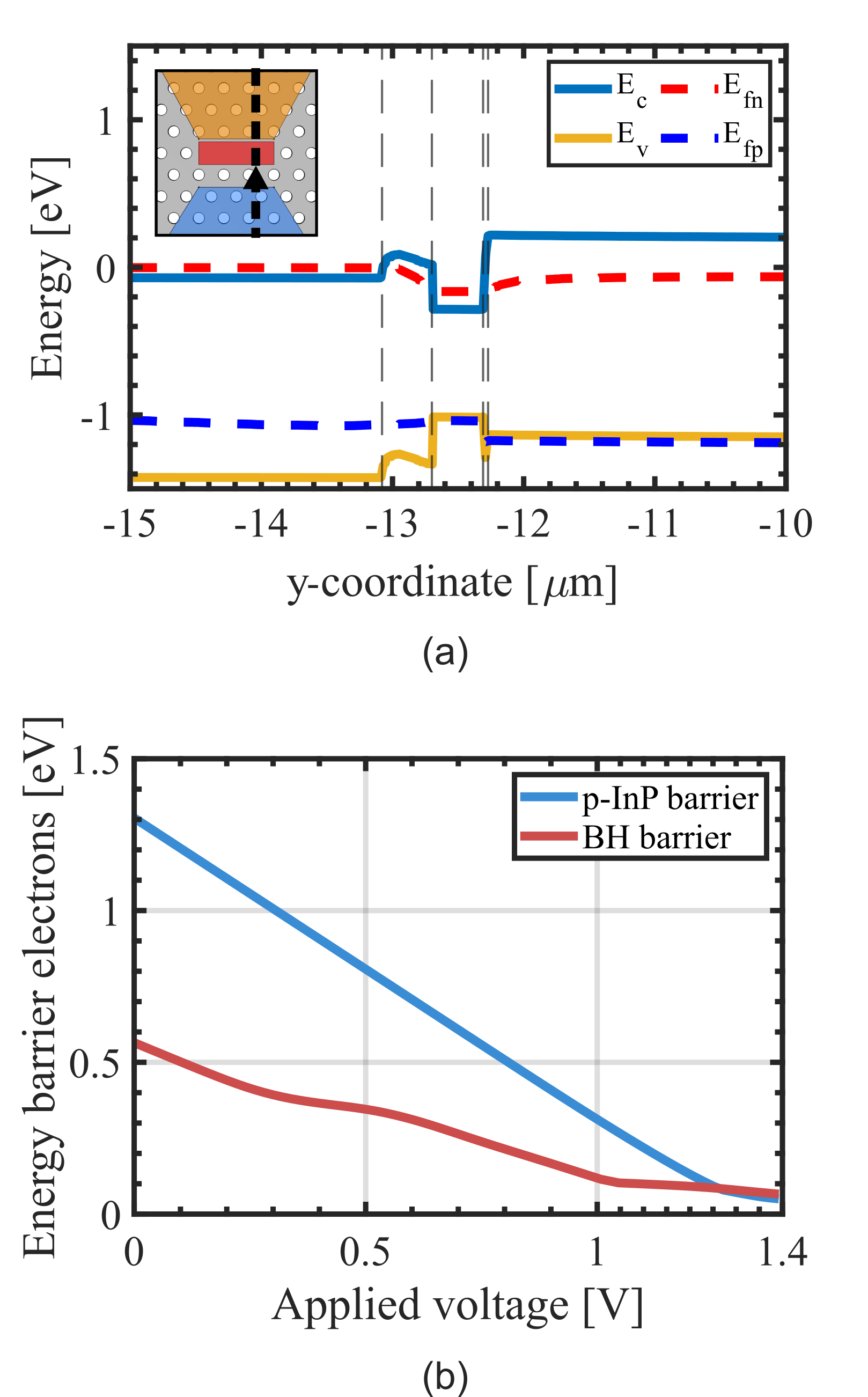}
    \caption{(a) Electronic band structures at $V_{\mathrm{bias}}$ = 1.2 V along the dashed lines in the inset. The arrow on the dashed line indicates the direction of increasing y-coordinate. (b) Energy barrier that needs to be overcome for electrons coming from the n-doped InP to reach the BH (red) and p-doped InP (blue) regions. \label{fig:bandstructure_poff1} }
\end{figure}

Figure \ref{fig:XY_IE_IQE_IV}(a) and \ref{fig:YZ_IE_IQE_IV} show the injection efficiency, $\eta_{\mathrm{inj}}$, internal quantum efficiency, $\eta_{\mathrm{IQE}}$, and total efficiency, $\eta_{\mathrm{tot}}$, as a function of the applied voltage for the XY and YZ simulations. From these results, we conclude that the device's performance is limited by the leakage paths in the XY plane and we shall therefore concentrate on the discussion of these. \\ \\ For the XY simulation, Figure \ref{fig:XY_IE_IQE_IV}(a) shows that the injection efficiency is initially close to 100$\%$ before it begins to drop off around a voltage of 0.9 V. This is caused by the formation of carrier leakage paths where electrons pass around the active region. As the voltage increases further, these leakage paths become more prominent, reducing the amount of carriers that reach the active region. \\ \\
To understand the trend in the internal quantum efficiency observed in Figure \ref{fig:XY_IE_IQE_IV}(a), it is useful to consider the recombination rates, cf. Figure \ref{fig:XY_IE_IQE_IV}(b). The different recombination rates, integrated over the active region area, S$_\mathrm{BH}$, (solid lines) and integrated over the InP area, S$_\mathrm{InP}$, (dashed lines), are shown versus the applied voltage. The surface SRH recombination is integrated over all etched surfaces and is combined with the bulk SRH recombination. Equation (S1)-(S4) in the Supporting Information show how the different recombination rates depend on the carrier density. At low voltages, non-radiative SRH recombination dominates as it is proportional to the carrier density, $N$, resulting in a low $\eta_{\mathrm{IQE}}$. When the voltage is increased, the spontaneous recombination rate increases and starts dominating due to its $N^2$ dependence, causing $\eta_{\mathrm{IQE}}$ to increase. As the applied voltage is increased further, the internal quantum efficiency decreases as the non-radiative Auger recombination rate ($\propto$ $N^3$) increases more rapidly. \\ \\ The onset of the stimulated recombination in the BH occurs around 1.07 V. At approximately 1.11 V, the lasing threshold is reached, causing the carrier density inside the BH to clamp. Additional carriers injected into the BH will now lead to stimulated recombination events, causing $\eta_{\mathrm{IQE}}$ to approach 100$\%$ as the voltage is further increased. Figure \ref{fig:XY_IE_IQE_IV}(b). For voltages below 0.9 V, the total efficiency follows the internal quantum efficiency as the injection efficiency is nearly 100$\%$ in this range, while it approximately follows the injection efficiency for higher voltages, where the internal quantum efficiency maintains a value greater than 80$\%$. \\ \\
\begin{figure*}
    \centering
    \includegraphics[width=0.75\linewidth]{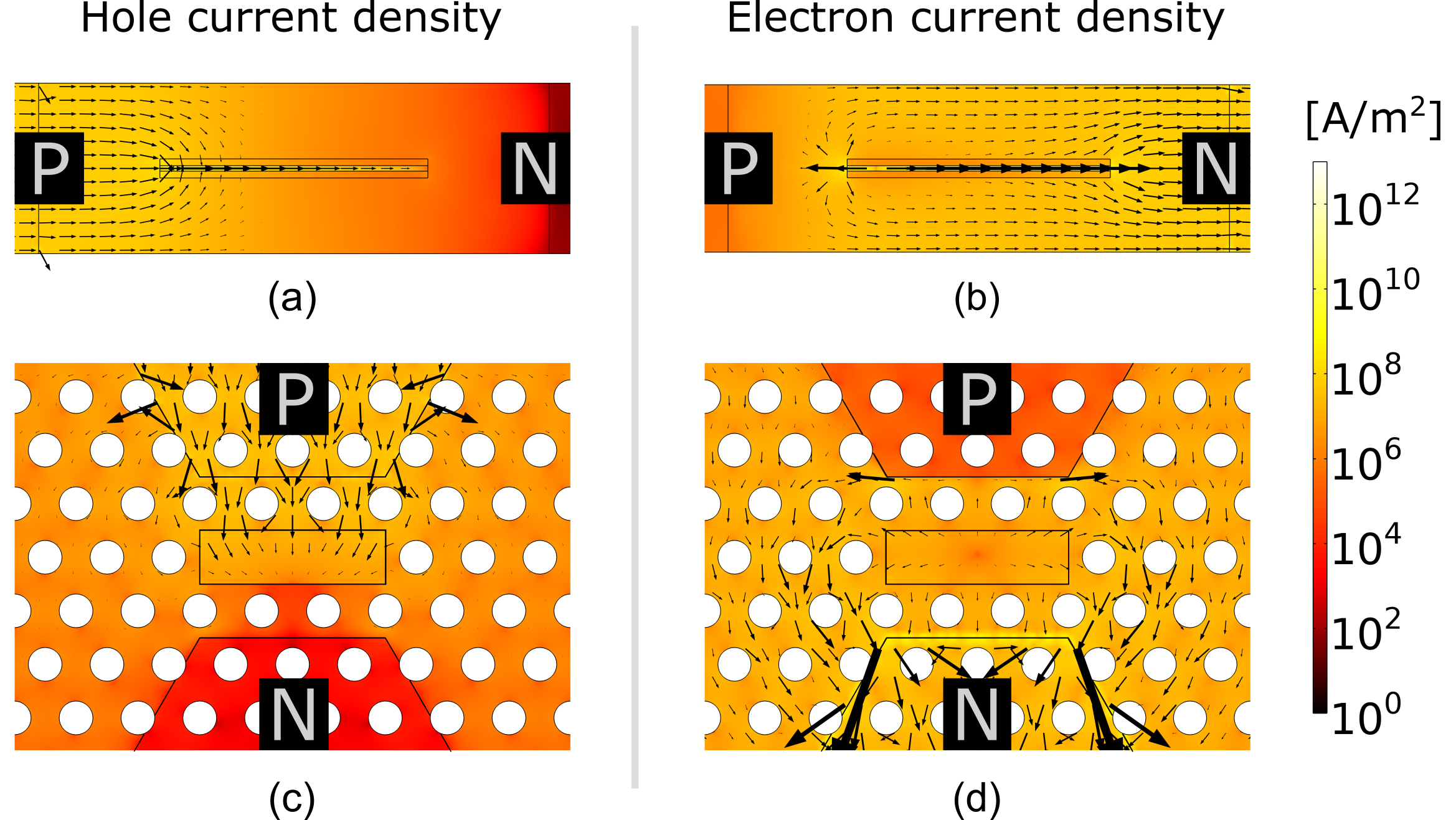}
    \caption{\label{fig:JnJp} 2D surface plots of the electron (J$_n$) and hole (J$_p$) current density in the YZ (top figures) and XY simulations (bottom figures) at a forward bias, V$_\mathrm{bias}$, equal to 1.3 V. Notice that the actual flow of electrons is opposite to the current. The colour represents the current density norm $||J|| = \sqrt{J_\mathrm{x}^2 + J_\mathrm{y}^2}$ while the arrows show the vector quantity (J$_\mathrm{x}$, J$_\mathrm{y}$) at specific grid points. The length of the arrows is scaled logarithmically in accordance with the surface plot scaling. The colour range of $||J||$ is consistent in all figures.}
\end{figure*}
Figure \ref{fig:YZ_IE_IQE_IV} shows the efficiencies versus applied voltage for the YZ simulation. Here, the injection efficiency remains close to 100$\%$ for all voltages in the considered range, and only starts to decrease slightly around 1.3 V, indicating that carrier leakage is limited in this cross-section. The internal quantum efficiency shows a similar trend to that in the XY simulation. However, the threshold of the laser, which matches closely with the dip in the internal quantum efficiency, occurs for a lower voltage than for the XY simulation. This can be explained by the consistently higher injection efficiency, causing the threshold gain to be reached at a lower applied voltage. \\ \\ 
In Figure \ref{fig:XY_IE_IQE_IV}(c), a log-log plot of the optical output power versus current and a semi-log plot of the internal quantum efficiency versus current are shown. The optical output power is calculated as \cite{Coldren_Corzine_ch2}:
\begin{equation}
    P_\mathrm{out} = \frac{N_\mathrm{p} h\nu V_\mathrm{p}}{\tau_\mathrm{p}}
\end{equation}
Here h is Planck's constant and $\nu$ the frequency of the emitted photons, set to 193.41 THz, corresponding to an emission wavelength of 1550 nm. The optimization of the outcoupling efficiency of the laser is a design problem related to the design of the cavity. Here, without loss of generality, it was set to 100$\%$ for simplicity reasons. A linear, spontaneous emission-dominated regime is present at low currents. Around the threshold current, a super-linear increase is observed while above threshold stimulated emission dominates, leading to a linear input-output relation.  A clear S-shaped input-output curve, characteristic of lasing, is obtained as a result. The threshold of the laser can be determined as the maximum of the first derivative of the log(L)-log(I) curve,\cite{Ning2013} and is indicated by the dashed line. The dip in the internal quantum efficiency closely aligns with the threshold current and will be used as a reference point when analyzing trends in the laser threshold. \\ \\
\noindent To further investigate the origin of the leakage paths, we plot the norm of the electron ($J_\mathrm{n}$) and hole current density ($J_\mathrm{p}$) in Figure \ref{fig:JnJp} at an applied voltage of 1.3 V for the XY and YZ simulation planes. The arrows depict the vector quantity (\textbf{$J_\mathrm{n/p,x}$}, \textbf{$J_\mathrm{n/p,y}$}) at specific grid points and their length is scaled logarithmically. Figure \ref{fig:JnJp}(a) and (c) show that the largest hole current density components are flowing into the BH in both simulation planes, with a limited fraction of the holes reaching the n-doped region. Figure \ref{fig:JnJp}(b) and (d) show, in contrast, that a significant fraction of the electrons flows around the BH. This effect is particularly prominent in the XY simulation, where the current density is highest in the photonic crystal region next to the BH (indicated by the large magnitude of the arrows). The presence of these unconventional leakage paths results in a reduced injection efficiency at higher voltages. Additional 2D surface plots of the electron and hole current densities for both simulation planes at voltages below 1.3 V are included in Note S5 of the Supporting Information.\\ \\ 
\begin{figure}[h]
    \centering  
    \includegraphics[width=0.9\linewidth]{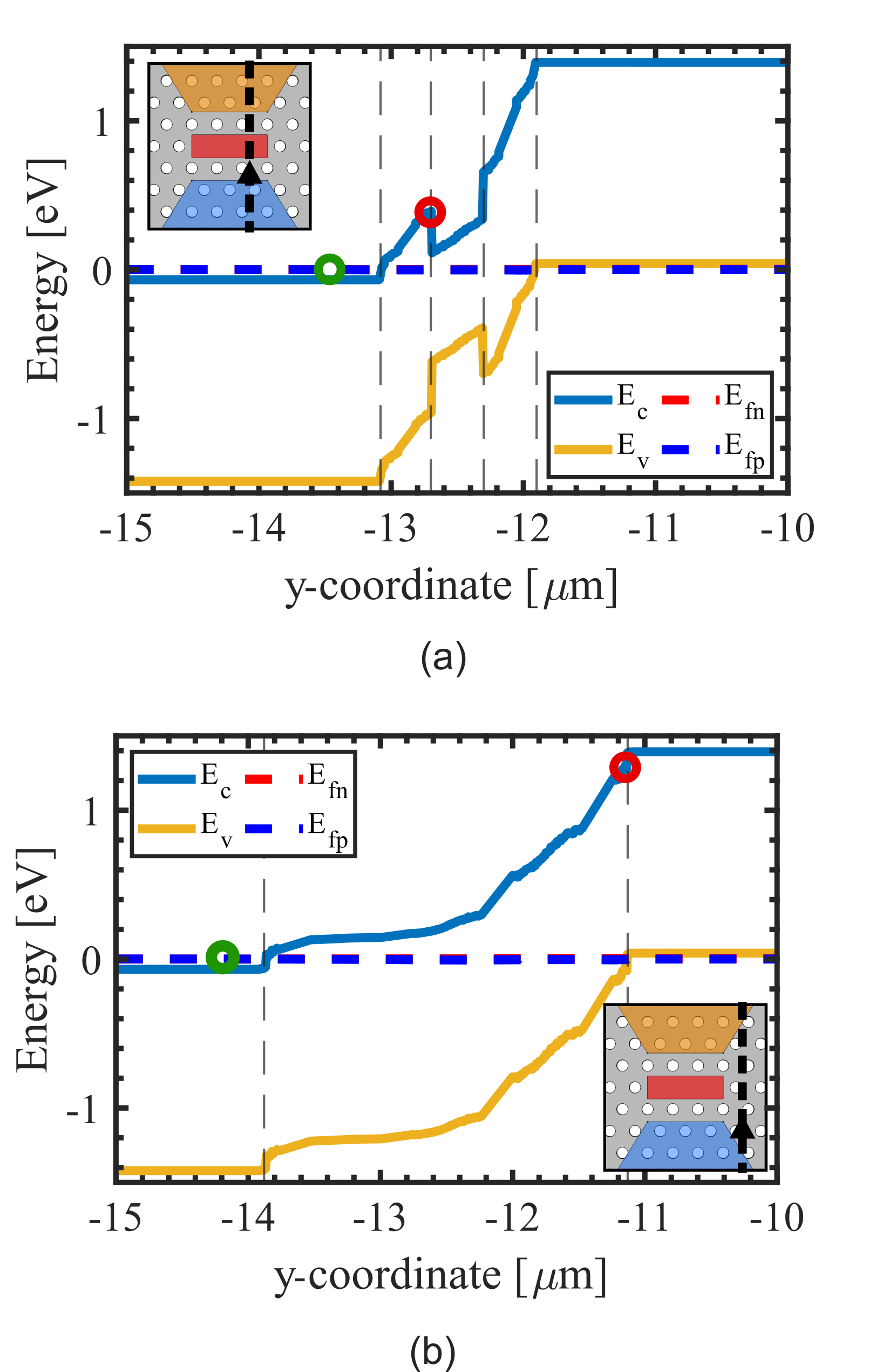}
    \caption{Electronic band structures at equilibrium ( $V_{\mathrm{bias}}$ = 0) along the dashed lines in the insets of (a) and (b). The arrow on the dashed line indicates the direction of increasing y-coordinate. The green and red circles indicate the energies used to calculate the energy barriers, shown in Figure \ref{fig:eBarriers}, experienced by electrons coming from the n-doped region. The vertical dashed lines indicate the interfaces between the p-, n-, i-InP, and BH regions. \label{fig:bandstructure}}
\end{figure}
To understand how these leakage paths arise, it proves useful to study the electronic band structure of the device. Figure \ref{fig:bandstructure}(a) and (b) show the band structure at equilibrium, $V_\mathrm{bias}$ = 0 V, along the dashed line shown in the inset of the figures. The vertical dashed lines in the band structures indicate the interface between the n-, i-, and p-doped InP regions and the BH region. Figure \ref{fig:bandstructure}(a) shows that the BH region acts as a potential well for both electrons and holes. As the device is in thermal equilibrium, the electron and hole quasi-Fermi levels, $E_\mathrm{fn}$ and $E_\mathrm{fp}$, are flat and degenerate. \\ \\ Using these band diagrams, one can determine the energy barrier electrons need to overcome to reach the BH and p-doped InP region. We define the BH barrier, experienced by electrons coming from the n-doped InP region, as the difference between the highest conduction band energy level to the left of the QW, indicated by the red circle, and the electron quasi-Fermi level in the n-doped InP region, indicated by the green circle. Similarly, from the band structure in Figure \ref{fig:bandstructure}(b), the energy barrier for electrons to reach the p-doped region can be determined as the difference between the maximum conduction band energy on the left of the p-doped region, indicated by the red circle, and the electron quasi-Fermi level in the n-doped InP region, indicated by the green circle. \\ \\ Figure \ref{fig:eBarriers} shows how the energy barriers vary as a function of the applied voltage. It shows that for voltages up to $\sim$1.1 V, it is energetically favourable for electrons to flow into the BH region as the energy barrier for this current path is lowest. However, for higher voltages, the energy barrier for electrons to reach the p-doped region becomes smaller, thus increasing the current flow through the PhC region, circumventing the BH active region. The energy barriers experienced by the holes show a similar trend with a crossing of both barriers at $\sim$1.15 V. A higher hole current density is therefore expected in the photonic crystal regions above this voltage. However, as seen in Figure \ref{fig:JnJp}(c) and (d), electrons, due to their high mobility, reach the p-doped InP region where they recombine with the holes. As a result, the hole current density in the photonic crystal is significantly reduced once it enters the i-InP region. This is signified by the reduction in the magnitude of the hole current density just outside the p-doped region. \\ \\ Insufficient injection of holes into the BH results in a net negative charge inside the BH, resulting in an additional electric field that increases the barrier for electrons to reach the BH. It is believed that this effect causes the BH barrier to decrease more slowly with voltage than the p-InP barrier. This can be observed in Figure \ref{fig:bandstructure_1_2V}, where the bandstructure along the dashed line in the inset is shown for $V_\mathrm{bias}$ = 1.2 V. Here, there is a higher barrier for electrons to enter the BH than for holes. In particular, the valence band is bent upward close to the BH on the p-side.  \\ \\ 
\begin{figure}[h]
    \centering
    \includegraphics[width=0.9\linewidth]{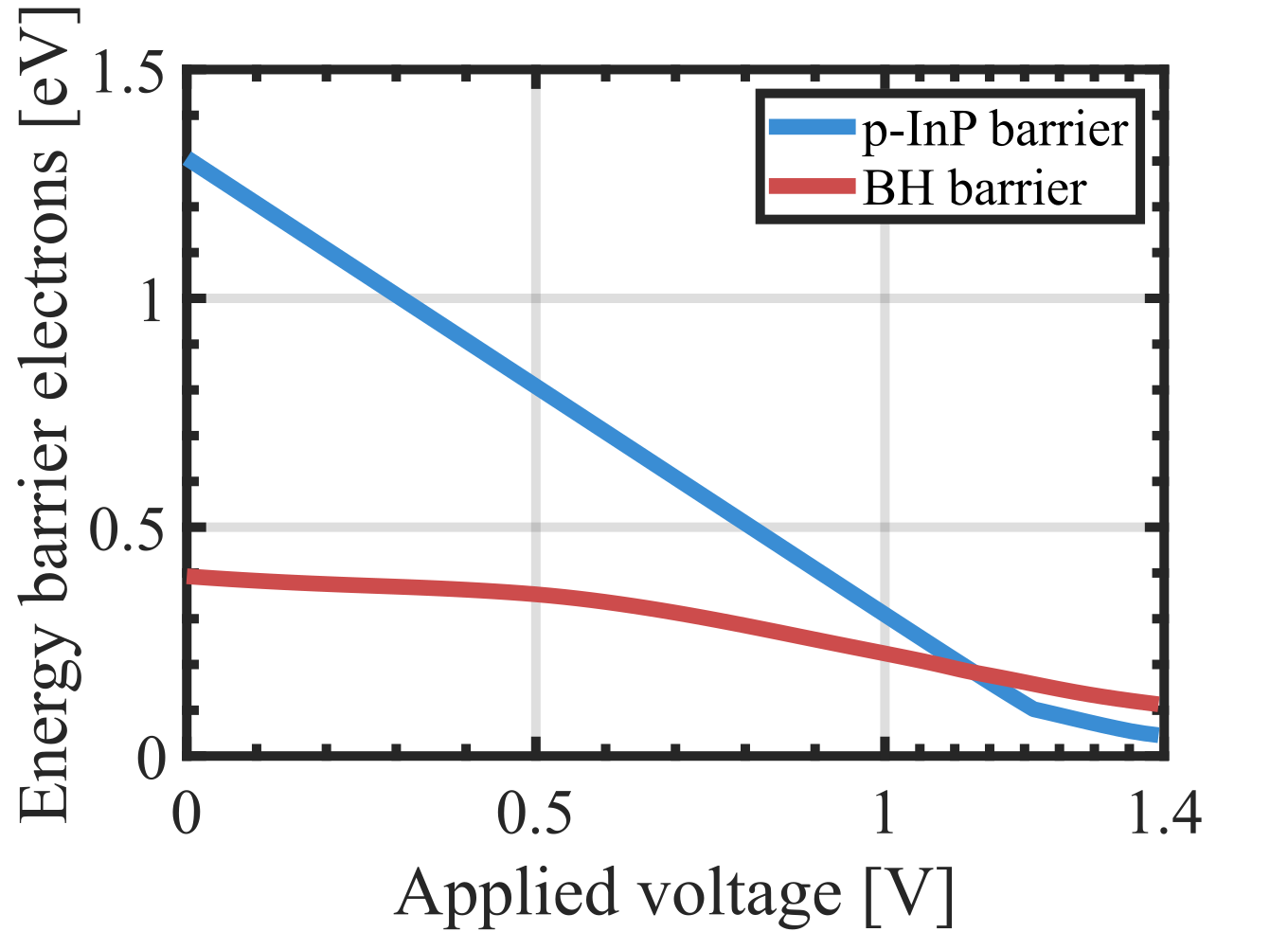}
    \caption{Energy barrier that needs to be overcome for electrons coming from the n-doped InP to reach the BH (red) and p-doped InP (blue) regions. The barrier is computed as the difference between the red and green circles in Figure \ref{fig:bandstructure}(a) for the BH barrier and Figure \ref{fig:bandstructure}(b) for the InP barrier. \label{fig:eBarriers}}
\end{figure}
\begin{figure}[h]
    \centering  
    \includegraphics[width=0.9\linewidth]{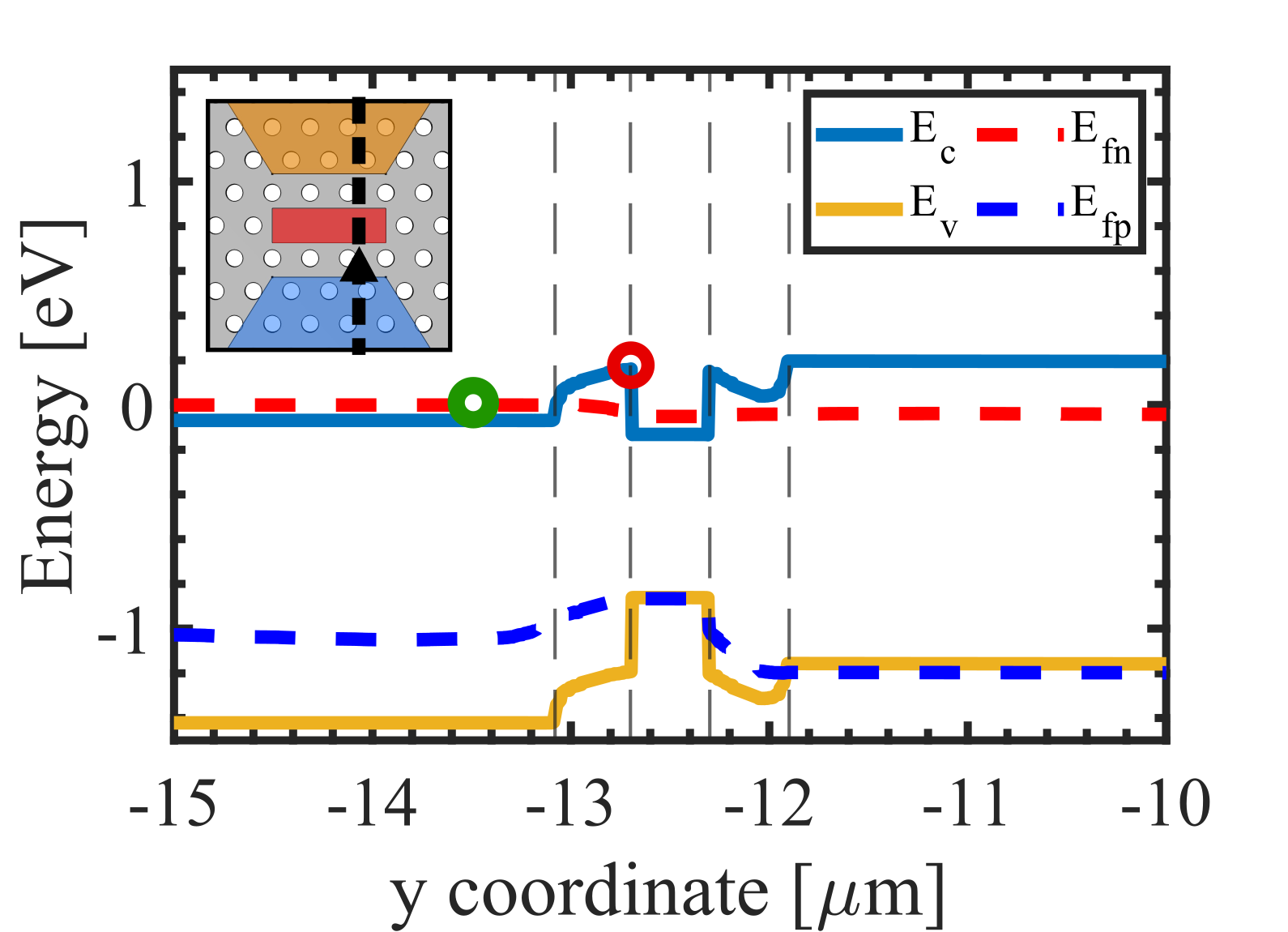}
    \caption{Electronic band structures at a forward bias $V_\mathrm{bias} = $ 1.2 V along the dashed lines in the insets. The arrow on the dashed line indicates the direction of increasing y-coordinate. The green and red circles indicate the energies used to calculate the energy barriers, shown in Figure \ref{fig:eBarriers}, experienced by electrons coming from the n-doped region. The vertical dashed lines indicate the interfaces between the p-, n-, i-InP and BH regions. \label{fig:bandstructure_1_2V}}
\end{figure}
\noindent Experimentally, these leakage paths manifest themselves in the observation of low injection efficiencies and enhanced spontaneous recombination in InP at the p-i interface.\cite{Dimopoulos2022} Figure \ref{fig:SpontRec}(a) shows a dimmed, top-view microscope image, captured with a Si camera, of an L12 line-defect laser under forward bias. Enhanced emission around the p-i interface signifies the recombination of electrons that reach the p-doped InP region. Figure \ref{fig:SpontRec}(b) shows a plot of the simulated spontaneous recombination rate in the XY-plane at an applied bias of 1.3 V. As expected, the spontaneous recombination rate is highest inside the BH where the electron and hole density are highest. The simulations further show good agreement with the experimental results, as a similar profile of enhanced spontaneous recombination is observed at the p-i interface. \\ \\
In the previous section, we mentioned that the observed low injection efficiency at high voltages is related to the low injection of holes into the BH region. Therefore, the offset of the p-doped region is a critical design parameter to improve the device's injection efficiency. In the following, we investigate the effect of the offset of the p-doped region on injection efficiency, internal quantum efficiency, and band structure of the laser to understand how the design can be optimized for injection efficiency. \\ \\ Figure \ref{fig:Schematics}(c) shows a schematic depicting the offset of the p-doped region from the BH region. The doping offset, $d_\mathrm{p}$, is given in units of $a_\mathrm{y}$ = $\frac{\sqrt{3}}{2}a$, where $a$ is the lattice constant of the photonic crystal. It is important to note that the p-doped region induces optical absorption losses, and there will therefore be a trade-off between optical cavity losses and injection efficiency. To account for the increased absorption losses that result from reducing the offset of the p-doped region, three-dimensional (3D) FDTD simulations of the optical field profile were performed to calculate the variation of the cavity Q-factor as a function of the offset of the p-doped region. This variation is incorporated in the model through the photon lifetime, which enters the photon density rate equation (see Equation \eqref{eq:Np}) and determines the threshold gain. The absorption losses from the p-doped region were characterized in Ref. \cite{Dimopoulos2022} and set equal to 120 cm$^{-1}$. \\ \\ Figure \ref{fig:pdoping}(a) shows the current-voltage (IV) characteristic of designs with varying offset of the p-doped region in XY simulations. The device has a p-i-n diode structure and therefore shows an IV characteristic similar to that of a diode. When forward biased, it conducts little current until the threshold is reached, after which the diode follows Shockley's diode equation. The result can be interpreted as an ideal diode in series with a parasitic resistance $R_\mathrm{p}$.\cite{ORTIZCONDE1999845} This leads to a linear current-voltage relationship above threshold with a slope equal to 1/$R_\mathrm{p}$. The earlier turn-on of the design with $d_\mathrm{p}$ = 0.6 $a_\mathrm{y}$ (cf. Figure \ref{fig:pdoping}(a)) can be attributed to the reduced energy barrier between the p-doped InP and the BH region as discussed below. This enables more efficient injection of holes into the BH at low voltages, leading to an increase in the current. \\ \\
Figure \ref{fig:pdoping}(b) shows the injection efficiency as a function of the applied voltage for four different p-doping offsets. As the offset of the p-doped region is decreased, the low mobility holes are injected more efficiently into the active region, increasing the injection efficiency. The effect on the laser's threshold voltage is seen in Figure \ref{fig:pdoping}(c), where the local minimum of the internal quantum efficiency, signifying the laser threshold, is shifted to lower voltages as the p-doped area is moved closer to the BH. To study the effect on the threshold current, Figure \ref{fig:pdoping}(d) shows the internal quantum efficiency as a function of the applied current. The trend in threshold current (indicated by the dip in $\eta_\mathrm{IQE}$) shows that there is a trade-off between the increased injection efficiency and increased doping-induced optical absorption losses as the offset of the p-doped region is varied. As a result, a minimum threshold current is found for the case where $d_\mathrm{p}$ = 1 $a_\mathrm{y}$. \\ \\
The effect of a reduced offset of the p-doped region is studied by calculating the BH and InP barrier for case $d_\mathrm{p}$ = 0.6 $a_\mathrm{y}$. A schematic of the design is shown in the inset. The electronic band structure, along the dashed line in the inset, at an applied voltage of 1.2 V, is shown in Figure \ref{fig:bandstructure_poff1}(a).  In the BH region, the hole quasi-Fermi level has penetrated into the valence band, indicating a high hole concentration. Simulations showed that the hole density is larger than the electron density in the BH region for all applied voltages. As a result, an additional field between the BH and n-doped region is created. This manifests itself in the conduction band having a negative slope in the i-InP region close to the BH. This lowers the energy barrier for electrons to reach the BH region and increases the injection efficiency of electrons into the BH. \\ \\ Figure \ref{fig:bandstructure_poff1}(b) shows the BH and p-InP barrier for electrons as a function of the applied voltage. The BH barrier crosses the p-InP barrier at an applied voltage of approximately 1.27 V, which is higher than observed in Figure \ref{fig:eBarriers}. Moreover, the difference in barrier height remains small, even for higher voltages. Therefore, the injection efficiency will be higher in this design as carrier leakage is limited. We note that this is only one design choice that can help alleviate carrier leakage. Other methods could include the current blocking trench introduced in Ref. \cite{Matsuo2019} or a 1D photonic crystal cavity, which reduces the presence of carrier leakage paths by design.

\section{Conclusion}
In this work, we investigate carrier transport in laterally injected 2D PhC lasers through finite-volume simulations. The model we developed allows us to study important electrical characteristics, such as the electrical injection efficiency and IV characteristics, while simultaneously capturing laser dynamics. We show that low injection efficiencies can be mainly attributed to electron leakage paths in the XY-plane. Due to the high mobility difference between electrons and holes, these leakage paths lead to enhanced spontaneous recombination at the p-i doping interface. We showed that an insufficient injection of holes into the active region increases the energy barrier for electrons to reach the BH. As a result, the injection efficiency decreases significantly at higher voltages. We further show that these results are consistent with experimental observations of low injection efficiencies and enhanced InP emission at the p-i interface. At last, we show that reducing the distance between the p-doped region and the active region results in a significant increase in injection efficiency. We believe that the model established in this paper and the insights gained on carrier transport in nanostructured semiconductor lasers are important for the development of a new generation of highly efficient and electrically-driven nanolasers.
    

    

\section*{Acknowledgements} \label{sec:acknowledgements}

This work was supported by the Danish National Research Foundation through NanoPhoton – Center for Nanophotonics (Grant No. DNRF147), The European Research Council (Grant No. 834410 FANO) and the Villum Fonden (Grant no. 42026 EXTREME).


\bibliography{main.bib}

\end{document}